\documentclass[preprint,showpacs,preprintnumbers,amsmath,amssymb]{revtex4}
\usepackage{graphicx}
\usepackage{dcolumn}
\usepackage{bm}
\def\gam5{\gamma_5}

\def\sigmaprim2{\sigma^{\prime 2}}
\def\piprim2{\vec{\pi}^{\prime 2}} 

\begin{document} 
\begin{center}
\vspace{5mm}
\Large{\bf The muon capture rate on hydrogen and the 
values of $g_A$ and $g_{\pi NN}$}

\vspace{8mm}
S. Pastore,    
F. Myhrer 
and 
K. Kubodera

\vspace{5mm} 
{\large Department of Physics and Astronomy,\\
\large University of South Carolina,\\ 
\large Columbia, SC 29208, USA }

\end{center}
\centerline{(\today)}
\vskip 1cm 

\vspace*{10mm}

\noindent
ABSTRACT ---  Motivated by the recent developments
in the determination of the experimental values of the nucleon axial-vector
coupling constant $g_A$ and the pion-nucleon coupling constant
$g_{\pi NN}$,  we carry out a heavy-baryon chiral perturbation calculation
of the hyperfine-singlet $\mu p$ capture rate $\Gamma_0$
to next-to-next-to-leading order (N$^2$LO), with the use of the latest values
of $g_A$ and $g_{\pi NN}$.  
The calculated N$^2$LO value is
$\Gamma_0^{\rm theor}(\mu^- p \to \nu_\mu n) =718 \pm 7 \,s^{-1}$, 
where the estimated N$^3$LO contribution dominates the error. 
This value is in excellent agreement with the experimental value
reported by the MuCap Collaboration.

\newpage

Muon capture on the proton has been the subject of
intensive experimental and theoretical investigations;
for reviews, see Refs.~\cite{Fearing2003,Kammel2010}.
Recently, the MuCap Collaboration succeeded in measuring,
to 1 \% precision, 
the rate $\Gamma_0$ of muon capture from the hyperfine-singlet state
of a $\mu p$ atom~\cite{MuCap2007}.
The reported experimental value is
\begin{eqnarray}
\Gamma_0^{\rm exp}(\mu^- p \to \nu_\mu n) 
&=& 
714.9 \pm 5.4 ({\rm stat}) \pm 5.1 ({\rm syst})\, {\rm sec}^{-1}
\, .
\label{eq:Gammaexp}
\end{eqnarray} 
Heavy-baryon chiral perturbation theory (HBChPT) 
provides a systematic framework for calculating $\Gamma_0^{\rm theor}$,
and a number of HBChPT-based calculations 
have been reported~\cite{Ando2000,Bernard2001,Udit2013}. 
HBChPT~\cite{bkm95,Bernard2009,Scherer}  
involves two perturbative expansions, 
one in terms of the expansion parameter $Q/\Lambda_\chi \ll 1$ 
and the other in terms of  $Q/m_N \ll 1$.
Here $Q$ is a typical four-momentum  transfer
involved in the reaction,
$m_N$ is the nucleon mass, and
$ \Lambda_\chi \!\simeq \!4\pi f_\pi\!\simeq
 \!1\,{\rm GeV}$ is the chiral scale.
In order for the theory to match 
the experimental precision of 1\%, 
one needs to incorporate higher order terms 
%
in the expansion in  $Q/\Lambda_\chi$ and $Q/m_N$.
In Ref.~\cite{Udit2013} (to be referred to as RMK),
Raha {\it et al.} evaluated $\Gamma_0^{\rm theor}$ including 
correction terms up to
next-to-next-to-leading order (N$^2$LO). 
They reported 
$\Gamma_0^{\rm theor}=710\!\times\!(1\pm0.007)\,{\rm sec}^{-1}$ which   
at N$^2$LO includes  radiative corrections and  finite proton size effect. 
The evaluation of $\Gamma_0^{\rm theor}$ 
in HBChPT at N$^2$LO 
involves several low-energy constants (LECs),
and the accuracy of the calculated value of $\Gamma_0^{\rm theor}$ 
at this order
depends on the precision with which these LECs are known. 
Additional uncertainties are due to the truncation at N$^2$LO 
of HBChPT expansion. 
The rate of convergence estimated from the leading order (LO), 
the next-to-leading order (NLO) 
and the N$^2$LO contributions to $\Gamma_0^{\rm theor}$ 
found in Refs.~\cite{Ando2000,Bernard2001,Udit2013}, 
indicates that N$^3$LO corrections would contribute 
at most $\sim 1$\%~\cite{Udit2013}. 
In the following we shall primarily concentrate 
on the uncertainties associated with the 
N$^2$LO evaluation of $\Gamma_0^{\rm theor}$. 
As emphasized in RMK, the above 0.7\% theoretical error
is dominated by the possible variations in the experimental values
of the nucleon axial-vector coupling constant, $g_A$, 
and the pion-nucleon coupling constant, $g_{\pi NN}$. 
This situation motivates us to pay particular attention
to recent highly noteworthy developments
regarding the experimental values of 
$g_A$~\cite{Mund2013,Mendenhall2013}
and $g_{\pi NN}$~\cite{Baru2011},
and to reexamine the value of $\Gamma_0^{\rm theor}$,
taking into account these developments.
The purpose of the present note is to report on such a study.

We first briefly summarize the treatment
of the LECs in RMK.
An N$^2$LO calculation of $\Gamma_0^{\rm theor}$
involves four LECs: $g_A$,  $\tilde{B}_2$, $\tilde{B}_3$, 
and $\tilde{B}_{10}$. 
$\tilde{B}_2$ is determined from the Goldberger-Treiman (GT) discrepancy 
\begin{eqnarray}
\Delta_{GT} \equiv \frac{2m_\pi^2}{(4\pi f_\pi)^2 g_A}\, \tilde{B}_2 = 
\frac{g_A\ m_N}{g_{\pi NN}\ f_\pi} -1\, ,  
\nonumber 
\end{eqnarray} 
while Refs.~\cite{bkm95,bfhm98} relate $\tilde{B}_3$ and $\tilde{B}_{10}$
to the nucleon mean squared axial radius $\langle r_A^2\rangle$ and
the nucleon isovector mean squared charge radius $\langle r_V^2\rangle$, 
respectively, via 
\begin{eqnarray}
\tilde{B}_3 &=& \frac{g_A}{2} \left( 4\pi f_\pi \right)^2 \frac{\langle r_A^2 \rangle }{3} \, ,
\nonumber \\ 
 \frac{1}{6}\langle r_V^2 \rangle &=& - \frac{2 \tilde{B}_{10}(\Lambda_\chi )}{(4\pi f_\pi)^2} 
-\frac{1+7g_A^2}{6(4\pi f_\pi)^2} - 
\frac{1+5g_A^2}{3(4\pi f_\pi)^2}\ln\left(\frac{m_\pi}{\Lambda_\chi}\right) \ . 
\nonumber
\end{eqnarray}
Since the term associated with $\tilde{B}_{10}$ 
gives only 
$\sim$0.1\% contribution to $\Gamma_0^{\rm theor}$,
and since $\langle r_V^2\rangle$ 
is relatively well known~\cite{Simon1980,Lorenz2012},
variations in $\Gamma_0^{\rm theor}$ due to the uncertainty in $\langle r_V^2\rangle$
can be safely ignored;
RMK used a fixed value, 
$\langle r_V^2\rangle^{1/2} = 0.765$~fm~\cite{Mergell96}. 
The terms associated with 
$\tilde{B}_2$ and $\tilde{B}_3$ 
give  $\sim 0.7\%$ and $\sim 1.9\%$ contribution to 
$\Gamma_0^{\rm theor}$, respectively,
implying a more pronounced sensitivity 
of $\Gamma_0^{\rm theor}$ to variations in the input parameters 
entering $\tilde{B}_2$ and $\tilde{B}_3$. 
As for the $\tilde{B}_3$ contribution,
RMK found that  $\sim 10\%$ variation in $\langle r_A^2\rangle^{1/2}$
(or equivalently, in the axial mass parameter $m_A$)
causes $\sim 0.3\%$ changes in $\Gamma_0^{theor}$,
which are not totally negligible; 
it is to be noted that the 10~\% variation 
is a rather ample allowance for the uncertainty in $\langle r_A^2\rangle^{1/2}$.
The  value of $g_{\pi NN}$, 
which affects $\tilde{B}_2$ via $\Delta_{GT}$,
was extracted from nucleon-nucleon scattering
and pion-nucleon scattering~\cite{Nijmegen1997,KH83,Ericson2002,Bugg2003}, 
but the resulting values show significant scatter. 
As an estimated range of variation in $g_{\pi NN}$,
RMK adopted $g_{\pi NN} \!=\! 13.044$---$13.40$, 
the smaller value taken from Ref.~\cite{Nijmegen1997}
and the larger value from Ref.~\cite{KH83}.
Variations in $g_{\pi NN}$ within this range lead to
$\sim$ 0.2 \% changes in $\Gamma_0^{\rm theor}$. 
For $g_A$, RMK employed as an estimate of its uncertainty
the difference between the PDG 2002 value 
and the  PDG 2012 value~\cite{PDG2002,PDG2010,PDG2012}.
Variations in $g_A$ within this range cause
$\sim$ 0.6 \% changes in $\Gamma_0^{\rm theor}$; 
these changes arise primarily from 
the overall multiplicative factor 
$(1\!+\!3 g_A^2)$ that enters the expression 
for $\Gamma_0^{\rm theor}$, 
and also from the contribution of the $\tilde{B}_2$ term.
The estimated theoretical uncertainty of 0.7 \%
in $\Gamma_0^{\rm theor}$ was obtained
by taking the quadratic sum of the above-mentioned individual errors.
It is noteworthy that the radiative corrections, 
which contribute about 2\,\%  to $\Gamma_0^{\rm theor}$~\cite{Sirlin2007},  
are well under control and do not affect the uncertainty 
in $\Gamma_0^{\rm theor}$; see Ref.~\cite{Udit2013} for details.

We now turn our attention to the latest experimental developments 
regarding $g_A$ and $g_{\pi NN}$.
Historically, the value of $g_A$ recommended by PDG 
has been steadily increasing, and the 2012 PDG value
is $g_A=1.2701\pm 0.0025$~\cite{PDG2012}.
Very recently, however, two groups~\cite{Mund2013,Mendenhall2013}  
reported the value $g_A \simeq 1.276$,
extracted from the measurement of the asymmetry parameter $A$ 
in neutron beta decay.
This new value is significantly larger than the 2012 PDG value. 
It is noteworthy that this new value of $g_A$ is consistent 
with the recently revised value of the neutron mean lifetime,
$\tau= 880.1\pm 1.1$ s (S=1.8)~\cite{PDG2012,Wietfeldt2011},
as discussed in Ref.~\cite{Mund2013}. 
Furthermore, Ivanov {\it et al.}~\cite{Ivanov2013} pointed out
the possibility that  these new values of $g_A$ and $\tau$ 
resolve the ``antineutrino flux anomaly", 
a lingering problem 
in the nuclear reactor neutrino-oscillation experiments.
Regarding the value of $g_{\pi NN}$, 
in a recent notable study~\cite{Baru2011},
Baru {\it et al.} improved 
the Goldberger-Miyazawa-Oehme sum rule analysis of   
Ericson {\it et al.}~\cite{Ericson2002},
and deduced the value,
$g_{\pi NN} = 13.116\pm0.092$.
It is worth emphasizing that Baru {\it et al.}~\cite{Baru2011}
used the most recent value for the $\pi N$ scattering length $a^+$,
which had been determined from the high-precision 
$\pi d$ atom data~\cite{Strauch2010}. 
These important developments
motivate us to re-evaluate $\Gamma_0^{\rm theor}$ at N$^2$LO 
with the use of the value of $g_A$ obtained 
in Refs.~\cite{Mund2013,Mendenhall2013},
and the value  of $g_{\pi NN}$ deduced 
in Ref.~\cite{Baru2011}. 
As will be discussed in the concluding paragraph, it is assumed here
that the electromagnetic effects have been removed from 
these two experimentally determined hadronic constants.

In calculating $\Gamma_0^{\rm theor}$, we use exactly
the same formalism and the input parameters 
as employed in RMK, \underline{except} the values of
$g_A$ and $g_{\pi NN}$;
as explained above, we adopt here
$g_A=1.2758 \pm 0.0016$~\cite{Mund2013,Mendenhall2013}, and 
$g_{\pi NN}= 13.116\pm0.092$~\cite{Baru2011}. 
To assess to what extent the uncertainties in $g_A$ and $g_{\pi NN}$
affect the precision in $\Gamma_0^{\rm theor}$,
we calculate $\Gamma_0^{\rm theor}$ for four cases.
In the first and second cases, $g_{\pi NN}$ is fixed at its central value 
$g_{\pi NN}= 13.116$, while $g_A$ is taken to be 
at the lower or upper end of the range 
within the experimental error.
In the third and fourth cases, $g_A$ is fixed at its central value,
$g_A= 1.2758$, while $g_{\pi NN}$ is assumed to be 
at the lower or upper end of the range 
within the experimental error.
Table~\ref{tab:predictions} shows the values of $\Gamma_0^{\rm theor}$
along with $\Delta_{GT}$
calculated for these four cases.
We emphasize that the results in this table comprise 
the radiative corrections and the finite proton-size effects,
as estimated in RMK.
Table~\ref{tab:predictions} indicates that the uncertainty in $g_A$ 
causes $\sim 0.2 \%$ variation  in $\Gamma_0^{\rm theor}$, 
and that the uncertainty in $g_{\pi NN}$ leads to $\sim 0.1\%$ variation.
To deduce the total uncertainty in $\Gamma_0^{\rm theor}$,
we recall that, according to RMK,
if one assigns 10 \% error to $\langle r_A^2\rangle^{1/2}$
(which is considered to be a rather generous error estimate),
it causes about  0.3~\% variations in $\Gamma_0^{\rm theor}$ at N$^2$LO.
By taking the squared sum of the errors 
that arise from $g_A$, $g_{\pi NN}$ and $\langle r_A^2\rangle^{1/2}$,
we arrive at 
\begin{equation} 
\Gamma_0^{\rm theor}({\rm N}^2{\rm LO}) =718 \times (1\pm 0.003)\,{\rm sec}^{-1}
\label{eq:GammaNew}
\end{equation} 
It is noteworthy that the new larger value for $g_A$~\cite{Mund2013,Mendenhall2013}  
increases the central value of $\Gamma_0^{\rm theor}$ by about 0.8 \%,
as compared with the result in RMK; 
this change arises primarily from the overall factor $(1\!+\!3 g_A^2)$
contained in the expression for $\Gamma_0^{\rm theor}$. 
It is also to be noted that the adoption of the new input 
for $g_A$ and $g_{\pi NN}$ significantly reduces the 
uncertainties in $\Gamma_0^{\rm theor}$ obtained in 
an N$^2$LO calculation.
Corrections entering at N$^3$LO are reasonably expected to produce 
at most a $\sim$ 1~\% contribution to $\Gamma_0^{\rm theor}$,
uncertainties that are within the present experimental precision. 
Since the 0.3 \% uncertainty that arises within an N$^2$LO calculation
is much smaller than that due to the possible N$^3$LO contributions,
it is reasonable to adopt the central value of  $\Gamma_0^{\rm theor}$
in Eq.(\ref{eq:GammaNew}) and attach $\sim$ 1~\% error to it:
$
\Gamma_0^{\rm theor}=718 \times (1\pm 0.01)\,{\rm sec}^{-1}
$

\begin{table}
\begin{center}
\caption{Capture rate, $\Gamma_0^{\rm theor}$,  
and Goldberger-Treiman discrepancy, $\Delta_{GT}$,
calculated with $g_A=1.2758 \pm 0.0016$~\cite{Mund2013,Mendenhall2013}, and 
$g_{\pi NN}= 13.116\pm0.092$~\cite{Baru2011}. $\Gamma_0^{\rm theor}$ is
evaluated to N$^2$LO, including radiative and proton finite-size corrections
as discussed in Ref.~\cite{Udit2013}. 
}
\label{tab:predictions}
\begin{tabular}{c|c|c|c}
\hline
$g_A$   &    $g_{\pi NN}$ &   $\Delta_{GT}$ 
&\,$\Gamma_0^{\rm theor}$ (s$^{-1}$) 
\\[1ex]
\hline
\hline
1.2774    &     13.116   & -0.011 
& 719.7  \\ 
1.2742    &     13.116   & -0.013   
& 716.9  \\ 
\hline
1.2758    &     13.208   & -0.019 
& 717.4 \\ 
1.2758    &     13.024   & -0.005 
& 719.2 \\ 
\hline 
\end{tabular}
\end{center}

\end{table}

\vspace{6mm}
To summarize, we have updated the HBChPT calculation 
of the hyperfine-singlet $\mu p$ capture rate $\Gamma_0^{theor}$
to N$^2$LO carried out in Ref.~\cite{Udit2013}, 
using the recently reported values 
of $g_A$ and $g_{\pi NN}$. 
We have assumed in this work that the coupling constants, $g_A$ and $g_{\pi NN}$, 
are pure hadronic constants. 
The electromagnetic corrections to, e.g. the 
asymmetry parameter, $A$, in polarized neutron beta decay which is 
used by Refs.~\cite{Mund2013,Mendenhall2013} to determine $g_A$, are 
known to be very small, e.g., Ref.~\cite{FK2004} finds 
radiative corrections to $g_A$ determined from $A$
to be 0.12\%. 
As to the value of $g_{\pi NN}$ the subtraction constant in the sum rule
has been extracted from pionic deuterium where, e.g., isospin violating effects 
are considered as well as QED effects. 
The hadronic cross sections entering the dispersion integrals  are 
also assumed to have been corrected for the
possible electromagnetic effects, see discussions in 
Ref.~\cite{Baru2011} and references therein. 
However, as shown in a highly illuminating paper 
by Gasser {\it et al.}~\cite{Gasser2003}, 
it is virtually impossible to extract pure hadronic values for, e.g.  
$g_A$ and $g_{\pi NN}$, from experimental data. 
With the use of
$g_A=1.2758 \pm 0.0016$~\cite{Mund2013,Mendenhall2013}, 
 and 
$g_{\pi NN}= 13.116\pm0.092$~\cite{Baru2011}, 
where we assume that the  errors quoted 
include residual electromagnetic effects, 
the theory favors a larger central value for $\Gamma_0^{theor}$ 
compared to the previous result~\cite{Udit2013}.  
In particular, our calculation
that includes radiative and proton finite-size corrections is
\begin{equation}
\Gamma_0^{\rm theor}(\mu^- p \to \nu_\mu n) =718 \pm 7\,
\,s^{-1}\, , 
\label{eq:GammaFinal}
\end{equation} 
where the error is dominated by the        
estimated N$^3$LO contributions. 
This new central value for $\Gamma_0^{\rm theor}$ 
is still in excellent agreement with the experimental value, 
Eq.~(\ref{eq:Gammaexp}), 
reported by the MuCap Collaboration.


\noindent
{\bf Acknowledgements}\\
This work is supported in part by  the National Science Foundation,
Grant No. PHY-1068305.


\end{document}